First-principles study on Small Polaron and Li diffusion in layered $LiCoO_2$

Seryung Ahn[1], Jiyeon Kim[2,3], Bongjae Kim[4], Sooran Kim[2*]

[1]Department of Physics, Kyungpook National University, Daegu 41566, South Korea

[2]Department of Physics Education, Kyungpook National University, Daegu 41566, South Korea

[3]The Center for High Energy Physics, Kyungpook National University, Daegu 41566, South Korea

[4] Department of Physics, Kunsan National University, Gunsan 54150, Korea

* Corresponding author: sooran@knu.ac.kr

**Abstract**

Li-ion conductivity is one of the essential properties that determine the performance of cathode materials for Li-ion batteries. Here, using the density functional theory, we investigate the polaron stability and its effect on the Li-ion diffusion in layered $LiCoO_2$ with different magnetic orderings. The localized $Co^{4+}$ polaron appears in the magnetic configurations and sets the Li-diffusion barrier of ~0.34 eV. The polaron also migrates in the opposite direction to the Li-diffusion direction. On the other hand, the polaron does not form in the non-magnetic structure, and the Li diffusion barrier without the polaron is 0.21 eV. Although the existence of the polaron increases the diffusion barrier, the magnetically ordered structures are more energetically stable during the migration than the non-magnetic case. Thus, our work advocates the hole polaron migration scenario for Li-ion diffusion. Moreover, we demonstrate that the strong electron correlation of Co ions plays an essential role in stabilizing the $Co^{4+}$ polaron.

**Introduction**

Lithium-ion batteries (LIB) have become key elements as a power source for current portable electronics.[1–3] Among various cathode materials for LIBs, layered lithium cobalt oxide, $LiCoO_2$ is a widely used cathode for rechargeable LIBs with its high energy density and long cyclability.[4,5] $LiCoO_2$ as a cathode material was originally suggested by Goodenough in 1980[6,7] and used in the first commercialized LIB by Sony in 1991.[2,8] Since then, $LiCoO_2$ is one of the most extensively studied cathode materials due to its broad use in modern technology revolution and rechargeable battery applications.[2,4,5]

$LiCoO_2$ is synthesized at high temperature (~800°C) in a hexagonal layered structure with the space group of $R\bar{3}m$ while it can be synthesized at low temperature (~400°C) in a cubic spinel structure.[9,10] The layered structure is formed by $Co^{3+}$-based edge-sharing $CoO_6$ octahedra separated from one another by the Li layer, which is suitable for Li-ion intercalation/deintercalation. Usually, this structure is termed as "O3 structure" where O represents the octahedral site for Li-ion, and the number indicates three types (AB, BC, and CA) of oxygen packing. Stoichiometric $LiCoO_2$ is in a non-magnetic ground state with low-spin $Co^{3+}$ ions well-explained by ligand field splitting picture:[11,12] Due to the strong ligand field, $Co^{3+}$ $3d$ orbitals split into two $e_g$ and fully filled three $t_{2g}$.[12] The $t_{2g}$ bands further split into $a_{1g}$ and $e_g'$ under trigonal distortion.[13]

When Li is deintercalated in $LiCoO_2$, previous experimental studies evidenced the existence of $Co^{4+}$ charge state in $Li_xCoO_2$ (x < 1).[13–16] The $Co^{4+}$ ion has five $3d$ electrons and is in a low-spin state (S=1/2) with a half-filled $a_{1g}$ orbital.[13] Ménétrier *et al.* reported the localized $Co^{4+}$ character with the $^7Li$ NMR study upon Li deintercalation.[14] Ito *et al.*, also claimed a mixture of $3d^6$ $Co^{3+}$ and $3d^5$ $Co^{4+}$ and possible magnetic phase transition at 175 K in $Li_xCoO_2$ using the temperature dependent magnetic susceptibility study.[15] The photoelectron spectroscopy investigation suggested the oxidation state change from $Co^{3+}$ to $Co^{4+}$ under the deintercalation of Li.[16] Furthermore, according to the recent soft X-ray spectroscopy by Mizokawa *et al*.,[13] $Co^{3+}$ with S=0 is oxidized to $Co^{4+}$ with S=1/2 as the Li-ions are removed. This $Co^{4+}$ state is explained by the polaron picture, where the local lattice distortion and oxygen 2*p* holes are formed around $Co^{4+}$. These oxygen holes flow back during the Li-ion flow so that they can compensate for the lattice distortion.[13]

Despite the experimental evidence showing localized $Co^{4+}$ or small polaron in $Li_xCoO_2$, many Li-ion conductivity studies using first-principles calculations generally do not consider this polaronic effect[17–21] and only a few papers investigated the polaron diffusion in $Li_xCoO_2$.[22–25] Most of the local density approximation (LDA) or generalized gradient approximation (GGA) level calculations allow to obtain the Li-ion diffusion barrier without $Co^{4+}$ polaron.[17,19,24] On the other hand, Hoang *et al*. intensively studied the defect effect in $LiCoO_2$ using hybrid functional and reported that the defect migration barriers for electron and hole polarons are 0.32 eV and 0.10 eV, respectively.[22,23] Moradabadi *et al*. presented that the Li-ion diffusion barrier with localized $Co^{4+}$ charge increases about ~0.2 eV within the GGA+*U* framework.[24] While the role of the polaron in $LiCoO_2$ is gaining attention, the detailed Li-ion diffusion paths along with polaron have not been reported so far to the best of our knowledge. Moreover, Co-ion is a representative 3*d*-transition metal (TM), which is known to show magnetism and strong correlation effect. The role of magnetism and correlation effect on the polaron formation as well as the diffusion process is an essential issue. Thus, it would be worth investigating the interplay of charge, spin, and lattice on Li-ion diffusion in layered $Li_xCoO_2$.

In this paper, we have explored the Li-ion/polaron diffusion, electronic properties, and polaron stability in $Li_xCoO_2$. Figure 1 illustrates the crystal structure of bulk $LiCoO_2$ and supercell structure for the delithiated $LiCoO_2$. We compare the total energy and Li-diffusion barrier depending on the magnetic ordering. The magnetic structures with the localized $Co^{4+}$ polaron are more stable than the non-magnetic one without the polaron during the whole diffusion path. It suggests that Li-ion diffusion along with the polaron is more plausible in the actual system than without. When Li-ion migrates with the polaron, the polaron migration path is opposite to the Li-ion one. Furthermore, we discuss the electronic structure of $Co^{4+}$ polaron and the stability of the polaron upon the strength of the electron-electron correlation.

**Computational details**

All density functional theory (DFT) calculations were carried out using the Vienna *Ab initio* Simulation Package (VASP), which implements the pseudopotential plane wave method.[26,27] The Perdew-Burke-Ernzrhof generalized-gradient approximation (PBE-GGA) was used for the exchange-correlation functional.[28] The PBE+$U$ calculations were performed to account for the correlated $d$ orbitals of Co-ions, whose effective on-site correlations, $U_{eff} = U - J$ are 3.4 eV[29] unless specified otherwise. The various $U_{eff}$ values (0.0, 1.0, 2.0, 3.0, 4.0, 5.0 eV) and hybrid functional[30] are also employed to investigate the effect of electronic correlation on polaron stability in detail. The Heyd-Scueria-Ernzerhof (HSE)06 was used for the hybrid functional with $\alpha$=0.25.[30] A plane-wave kinetic energy cutoff is 650 eV for all calculations.

We fully relaxed the hexagonal $LiCoO_2$ with $R\bar{3}m$ space group and generated 4x4x1 supercells with two Li vacancies, corresponding to 190 atoms in total as shown in Fig. 1. The composition of the supercell with Li divacancy is $Li_{0.96}CoO_2$. We considered both the spin-polarized and non-spin-polarized calculations for the 4x4x1 supercells. The atomic positions of the 4x4x1 supercells were optimized until the residual forces were less than 0.01 eV/Å. The nudged elastic band (NEB)[31,32] method was used to calculate the Li-ion migration barriers with Li double vacancies. The NEB calculations were converged until the residual force was less than 0.05 eV/Å. The *k*-points for the 4x4x1 supercell were 2x2x2 and 1x1x1 for PBE+*U* and HSE06 calculations, respectively.

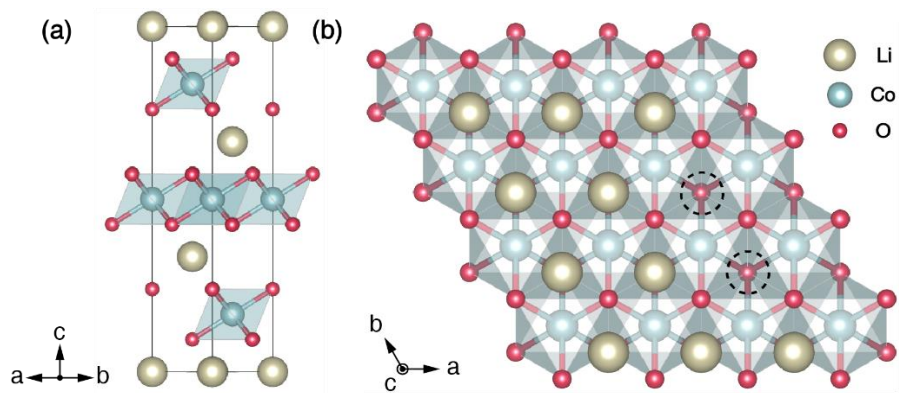

**Fig. 1** (a) Crystal structures of bulk layered $LiCoO_2$ and (b) top view of 4x4x1 supercell of $LiCoO_2$ with Li divacancy. The empty circles indicate the position of Li vacancies.

**Results and Discussion**

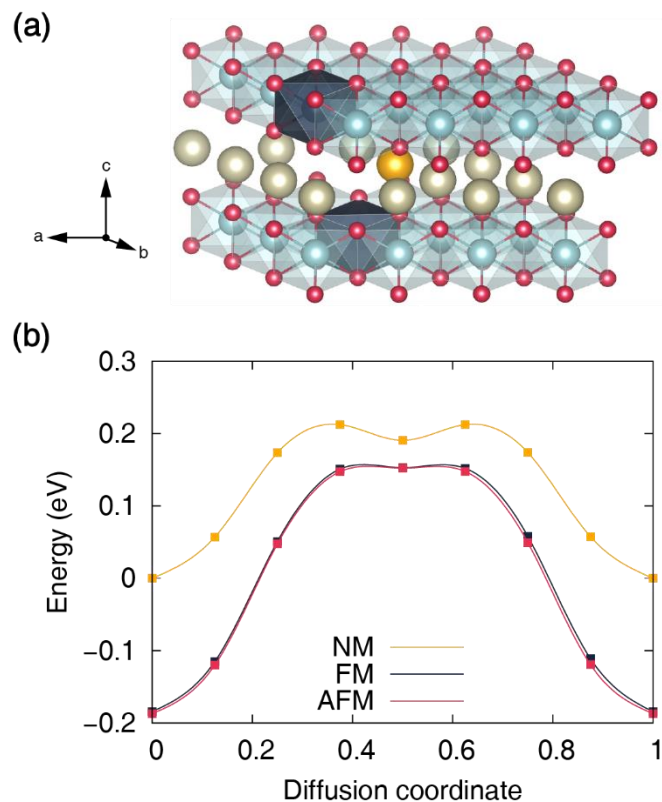

**Fig. 2** (a) Polaron formation in LiCoO$_2$ with Li divacancy. The yellow sphere and darker blue octahedra indicate the moving Li and Co$^{4+}$ positions, respectively. (b) Comparison of Li diffusion barriers with and without polaron in different magnetic orderings. Polaron does not localize in non-magnetic (NM) case while the polaron appears in ferro/antiferromagnetic (FM/AFM) structures.

To explore the polaron formation from the localized Co$^{4+}$, we relaxed the structures having Li divacancy with different magnetic orderings: non-magnet (NM), ferromagnet (FM), and antiferromagnet (AFM). The spins of two Co$^{4+}$ in the FM and AFM structures are in the same and opposite directions, respectively. We found that the polaron forms in magnetic cases only while it does not appear in the NM case. The average lengths of Co$^{3+}$-O bonds in the CoO$_6$ octahedra were 1.914 Å and 1.927 Å in non-magnetic and magnetic configurations, while the $Co^{4+}$-O bond in magnetic structures is 1.894 Å. The contracted TM-O bonds indicate the hole polaron formation. Since the size of the lattice distortion by the localized hole is confined to local $Co^{4+}O_6$ octahedra, this type of polaron can correspond to a small polaron.[33,34] Figure 2(a) illustrates the polaron positions with divacancy: two $Co^{4+}$ polarons form and localize near Li vacancies. $Co^{3+}$ is in $3d^6$ electronic configuration with a low spin state, thus only $Co^{4+}$ shows significant magnetic moments of 1.000 $\mu_B$ with a low spin state in the magnetic structures. The total energy of magnetic cases is ~4 meV/f.u. lower than that of the NM structure, indicating polaron stability. Figure 2 shows the energy difference per 4x4x1 supercell with a Li-ion migration. The energy of FM is only 0.04 meV/f.u. higher than AFM, which suggests the formation of the polaron is local and not very relevant to the magnetic exchanges. The stable polaron from the localized hole in $Co^{4+}O_6$ in $LiCoO_2$ with vacancies is consistent in the previous experiments.[13–16]

Once we identified the polaronic ground of the system, we performed NEB calculations to investigate the effect of polaron on the Li diffusion barrier. Previously, two Li diffusion paths are suggested in layered $Li_xCoO_2$: one is a single vacancy mechanism through oxygen dumbbell hopping and another is a divacancy mechanism through tetrahedral site hopping.[17,18,35,36] Because the migration barrier of ~0.2 eV via the divacancy mechanism[17–19,22] is much lower than that of ~0.4-0.8 eV through the single vacancy mechanism[17,18,20–22], lithium diffusion is expected to be predominantly through the divacancy mechanism.[17,18,36] Therefore, we have focused the divacancy model for the Li migration in this work.

Figure 2(b) shows the calculated migration barriers with different magnetic orderings. The energy barriers are 0.212, 0.338, and 0.339 eV for NM, FM, and AFM structures, respectively. The Li diffusion barrier of the NM case without the $Co^{4+}$ polaron agrees well with the previous

Li diffusion barrier of ~0.2 eV via the divacancy mechanism.[17–19,22] When Li-ion diffuses with polaron (FM/AFM cases), the diffusion barrier itself increases about ~130 meV, implying that the polaron deteriorates the Li diffusion. In the diffusion coordinates, however, the energy of magnetic structures (FM/AFM) is always lower than that of the NM structure despite a higher diffusion barrier. Therefore, even though the formation of polaron raises the barrier of Li migration, our calculations suggest that Li diffusion involving the polaron migration is energetically favored in $Li_xCoO_2$. In addition, the different magnetic ordering does not significantly alter the diffusion barrier as shown in the FM and AFM cases of Fig. 2(b).

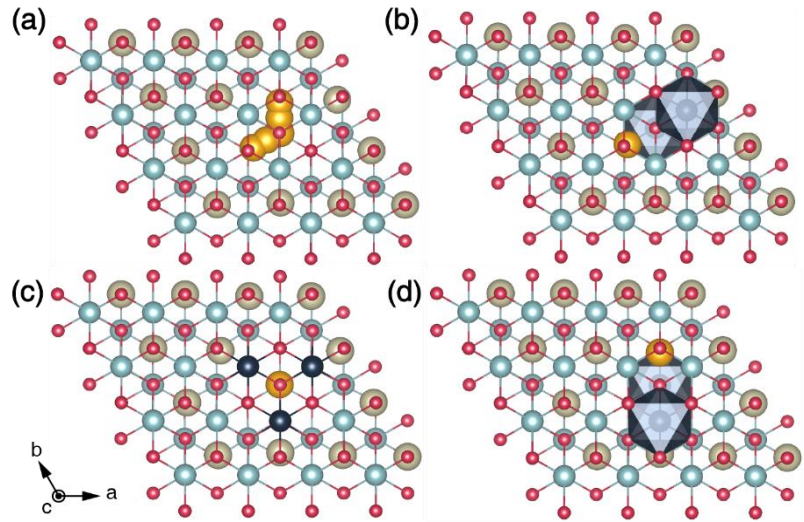

**Fig. 3** Li diffusion paths in divacancy mechanism without and with polaron migration. (a) Li migration in the NM structure (b-d) Li and polaron migration in the FM structure. (b), (c) and (d) illustrate the initial, middle, and final atomic configurations, respectively. The transparent polyhedral represents the $CoO_6$ octahedral with low spin $Co^{4+}$ polaron. The darker blue spheres in (c) indicate Co-ions with a magnetization of 0.148-0.177 $\mu_B$, which is smaller than that of $Co^{4+}$ but larger than that of low spin $Co^{3+}$.

Figure 3 illustrates the Li diffusion pathways in the NM (without polaron) and FM (with polaron) structures. For the divacancy mechanism[17,18,35,36], the Li-ion migrates from one octahedral site to a neighboring octahedral site through the tetrahedral site as shown in Fig. 3(a). On the other hand, when both Li-ion and $Co^{4+}$ polaron diffuse, the diffusion directions are opposite. There are two polarons near Li-vacancies: one is above the Li layer with the vacancy, and another is below the Li layer. As shown in Fig. 3(b) and (d), when Li-ion migrates upwards in Fig. 3(b-d), one polaron migrates downwards. Another polaron below migrating Li-ion keeps its position and does not migrate. These opposite directions of Li/polaron diffusions agree well with Li-ion flow with the oxygen hole surrounding $Co^{4+}$ backflow.[13]

When Li-ion is at the tetrahedral site in the diffusion path, the polarons are delocalized as in Fig. 3(c): instead of two $Co^{4+}$ polarons with the magnetization of 1.000 $\mu_B$, three $Co^{3.x+}$ appear with the magnetization of 0.148, 0.159 and 0.177 $\mu_B$. In addition, the Li and polaron diffusions in the AFM ordering are the same as the FM case, suggesting the energy scale involved in the polaron formation and lattices are much larger than the magnetic exchanges.

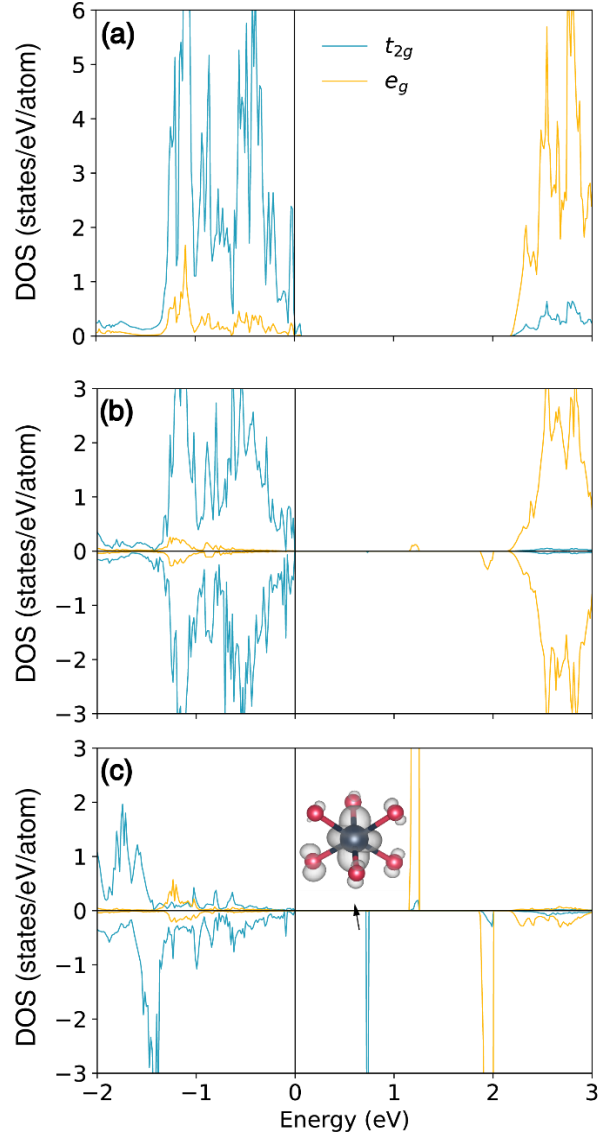

**Fig. 4** Projected density of states (PDOS) of 3$d$ orbitals of (a) Co-ion in the NM structure. (b) Co$^{3+}$, and (c) Co$^{4+}$ in the FM structure. (Inset) Charge density isosurface of CoO$_6$ octahedral $t_{2g}$ hole band.

Now, we look further into electronic structures. We have calculated the projected density of states (PDOS) of the Co 3$d$ orbitals in the NM and FM structures as in Fig. 4. For the NM structure, $t_{2g}$ bands are mostly occupied while $e_g$ bands are unoccupied, which indicate the low spin state of Co-ion. The DOS of Co ion in the NM case is almost the same as that of Co$^{3+}$ in the FM case except for a slight hole state of $t_{2g}$ orbital near the Fermi level. These $t_{2g}$ hole bands originate from the Li divacancy oxidizing Co-ions overall, which results in the metallic phase.

The FM structure shows a clear separation between $Co^{3+}$ (Fig. 4(b)) and $Co^{4+}$ (Fig. 4(c)). The six electrons of $Co^{3+}$ fully occupy the $t_{2g}$ triplet with the low spin state while five electrons of $Co^{4+}$ generate hole bands in $t_{2g}$ bands with the magnetic moment of 1.000 $\mu_B$. The FM structure exhibits the insulating phase with a band gap of 0.72 eV. The band gap arises from the gap between $t_{2g}$ orbitals of $Co^{4+}$ ion as shown in Fig. 4(c). In the case of $Co^{4+}$, the unoccupied $e_g$ exhibits two peaks and part of $t_{2g}$ becomes one unoccupied peak. The inset of Fig. 4(c) illustrates the charge density of the unoccupied $Co^{4+}$ $t_{2g}$ orbital. The charge density indicates that the unpaired electrons are located at the $a_{1g}$ orbital, which is consistent with the previous studies.[13,37] Furthermore, we performed the Bader charge analysis on oxygens around $Co^{3+}$ and $Co^{4+}$, and the average charges of oxygen atoms are 7.064 and 6.982, respectively. The smaller oxygen Bader charge around $Co^{4+}$ than that of $Co^{3+}$ indicates the oxygen hole bounded around $Co^{4+}$, which agrees well with the previous experiments.[13,38] The charge density in Fig. 4(c) also shows the oxygen hole around $Co^{4+}$. In addition, the electronic structure of the AFM structure is similar to that of the FM case.

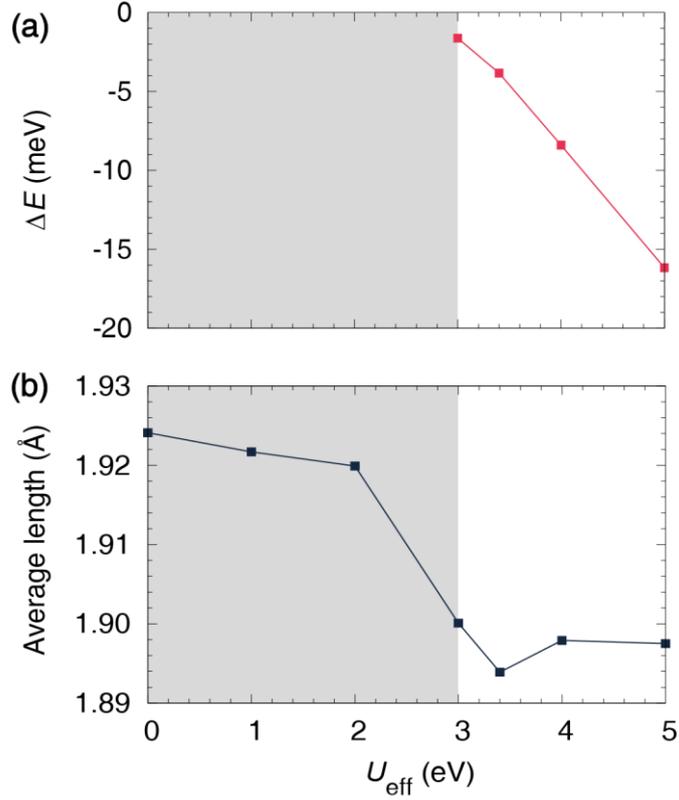

**Fig. 5** Polaron stability depending on the electron correlation. (a) The energy difference between polaron and non-polaron structures per formula unit and (b) Average Co-O bond length of $CoO_6$ as a function of $U_{eff}$. The gray shaded area represents where the $Co^{4+}$ polaron is not stabilized.

The formation and behaviors of polarons, especially, at the TM-ion sites, are known to largely depend on the degree of the localization of the electron charges. Hence, the strength of the electron-electron correlation, the so-called $U$, plays an important role in polaron stability. For this, we have studied the role of the $U$ within the DFT+$U$ scheme on the stability of the polaronic ground by calculating the energy difference, $\Delta E$ between polaron and non-polaron structures and average Co-O bond length of $Co^{4+}$ octahedra as a function of $U_{eff}$ from 0 to 5 eV. Figure 5(a) shows that the $Co^{4+}$ polaron starts to be stable when $U_{eff}$ is larger than 3 eV and the magnitude of $\Delta E$ increases as the $U_{\text{eff}}$ increases. This shows that a sizable correlation effect is essential for polaron formation, and the critical $U_{\text{eff}}$ for the polaron formation is 3eV. Note that the usual $U_{\text{eff}}$ values employed for $Co^{3-4+}$ is about ~3-6 eV.[19,30,39,40] We further performed the hybrid HSE06 functional calculation, which is known to include the longer-range Hartree-Fock exchange and effective for the description of the polaron. We found that the magnitude of $\Delta E$ becomes larger with the hybrid functional whose $\Delta E$ is -22.9 meV/f.u.. The HSE calculations also suggest the well-stabilized small polaron in $Li_xCoO_2$.

Figure 5(b) shows the average length of Co-O in the octahedra depending on the $U_{\text{eff}}$. When $U_{\text{eff}}$ is larger than 3eV, the average length is significantly reduced, which indicates the polaron formation. Although the magnitude of $\Delta E$ becomes larger with increasing $U_{\text{eff}}$, the average length does not notably change after $U_{\text{eff}}$=3eV. It suggests that the increased magnitude of $\Delta E$ with $U_{\text{eff}}$ > 3eV originates from the electronic part rather than lattice deformation.

**Conclusions**

In conclusion, we demonstrate that the interplay of magnetic ordering and strong Coulomb correlation can affect the polaron formation and Li-ion diffusion using first-principles calculations. The $Co^{4+}$ polaron only appears with the FM/AFM magnetic structure while the NM one does not exhibit the polaron. We have found that the existence of polaron increases the diffusion barrier, but decreases the total energy compared to the energy without the polaron formation. Therefore, we expect that the actual Li migration occurs with the polaron. While Li-ion diffuses, the polaron also migrates whose direction is opposite to the direction of Li diffusion. Furthermore, we show that the strong electron correlation of Co-$d$ orbital stabilizes the polaron formation. We hope that this study can stimulate the detailed investigation of the alkali-ion diffusion under various conditions such as magnetic ordering, strong electronic correlation, and lattice deformation for Li/Na-ion cathode materials.

## Conflicts of interest

There are no conflicts to declare

## Acknowledgements

We thank the fruitful discussion with Dr. Kyung-Tae Ko. This work was supported by the National Research Foundation of Korea (NRF) (Grant No. 2022R1F1A1063011, 2021R1C1C1007017), KISTI Supercomputing Center (Project No. KSC-2021-CRE-0495, KSC-2021-CRE-0605). S.K and B.K thank the hospitality of the PCS at IBS, Daejeon, Korea (IBS-R024-D1). BK acknowledges support from the Korea Institute of Energy Technology Evaluation and Planning (KETEP) grant funded by the Korea government (MOTIE) (20224000000220, Jeonbuk Regional Energy Cluster Training of human resources).